\documentclass[aps,prb,twocolumn,groupedaddress,showpacs]{revtex4}
\usepackage{graphicx}
\usepackage{latexsym}

\begin{document}

\title{Superconductivity in the insulating phase above the field-tuned superconductor-insulator transition in disordered indium oxide films}
\author{Myles Steiner}

\author{Aharon Kapitulnik}

\affiliation{Departments of Applied Physics and Physics, Stanford University,
Stanford, CA 94305}

\date{\today}

\begin{abstract} 
We study the insulating phase of disordered indium oxide films that undergo a field-tuned superconductor-insulator transition. The transport measurements in a perpendicular magnetic field show distinct regimes of strongly fluctuating order-parameter amplitude and phase, and reveal a wide range of insulator strength in samples with differing disorder, despite the similarity in behavior near the SIT. We characterize the strength of the insulating phase and compare it to the superconducting strength. We find that the films do not return to the expected normal state even at high perpendicular magnetic fields where all pairs should be broken, suggesting the remaining presence of superconductivity at high fields. 
\end{abstract}

\pacs{74.72.-h, 74.78.-w, 74.40.+k}

\maketitle

\section{Introduction}

The physics behind the superconductor-insulator quantum phase transition (SIT) in disordered films has been the subject of numerous studies. \cite{goldman1}  By tuning an external parameter such as the disorder or the magnetic field, the ground state of the system changes continuously from a superconductor, characterized by a resistance that tends to zero as temperature is lowered, to an insulator characterized by a diverging resistance. Experimentally, the magnetic field tuned transition is simplest to study as each experiment requires only one sample realizing a fixed amount of disorder, and the transition occurs at a particular field $H_c < H_{c2}(0)$ (we discuss the meaning of $H_{c2}$ in section \ref{sechc2}).   For $H < H_c$  isotherms of lower temperature show a tendency towards vanishing resistance, while for $H > H_c$ a resistance that increases with lower temperatures marks the emergence of an insulating phase.  While recent studies \cite{kap1} of the field-tuned transition suggest that dissipation effects may be important, revealing novel metallic phases that seem to persist even as $T\rightarrow 0$, the general features of the SIT can still be recovered if the right temperature window is used.  In this case one can use  the so-called ``dirty boson" model \cite{mpaf} to analyze the transition near $H_c$.  Indeed such a procedure was used successfully in many cases, \cite{hebard1,kap2,mk1,markovic} revealing a scaling behavior of the magnetoresistance.  In the presence of dissipation ``true superconductivity" in which the resistance is zero has been shown to exist at fields much lower than $H_c$ in amorphous-MoGe films. \cite{mk2}  The insulating phase has been shown to also have novel features such as a possible crossover towards Fermi system behavior, \cite{hebard2} leveling of the resistance at low temperatures, \cite{mk1} and possible existence of vortices \cite{markovic1} above the SIT, as well as possible activated behavior for highly disordered films. \cite{sambandamurthy1}  However, no systematic study of this phase and in particular its evolution with disorder has been done to explore  these features.

In this paper we present a set of measurements on indium oxide films that display a wide spectrum of insulator strength.  While the normal state resistance of the films just above the transition is in the range of 2.5 - 8 k$\Omega/\Box$, their insulating tendency spans the range from  logarithmic in temperature (``weak insulator") to an activated behavior with an activation temperature of order $T_{c0}$ (``strong insulator"). The films used for this study show a homogeneous amorphous structure as indicated by the study of their microstructure and their amplitude-dominated resistive transitions.  For all these films we show that at low temperatures and upon increasing the perpendicular magnetic field, the films first show the signatures of a SIT, followed by a distinct Bose-insulator behavior in which the resistance increases with field, reaching a peak resistance at a field of order of $H_{c2}(0)$.  At higher fields the resistance decreases again, first rapidly and then more gradually, suggesting that superconducting pairs persist to fields as high as three times the peak field. \cite{markovic}  This result may have implications to other systems that exhibit a similar SIT, and possibly to the high-$T_c$ cuprate superconductors.\cite{myles1}

\section{Experimental}

Indium oxide (InOx) is an amorphous low-carrier-density superconductor ($n\sim10^{20} - 10^{21}$ carriers/cm$^3$) \cite{hebard,kowal} that was used in previous studies of the superconductor-insulator transition (SIT). \cite{hebard1} While it may have been argued in the past that InOx films are not homogeneous, a more recent fabrication method introduced by Kowal and Ovadyahu \cite{kowal} showed that InOx can be made non-granular by a combination of e-beam evaporation and annealing at a relatively low temperature. Transmission electron micrographs of these films were shown to be completely amorphous, while comparison with electron diffraction patterns from pure indium films ruled out the presence of indium crystallites as small as $\sim10\AA$ which were observed in films prepared by other methods. \cite{hebard}  Following Kowal and Ovadyahu we prepared our films by electron beam evaporation of sintered In$_2$O$_3$ onto an acid-cleaned silicon nitride substrate. Control of the films' normal state resistance, superconducting transition temperature, and the overall superconductor-to-insulator transition are adjusted by adding oxygen during growth and by the subsequent careful annealing of the samples. An argon ion etch was used to pattern the films into Hall bars 100 $\mu$m wide and up to 1000 $\mu$m in length, with narrow voltage probes on each side. Throughout the preparation we were careful to keep the temperature below $60^oC$ to avoid recrystallization. After evaporating Ti-Au contact pads, the films were annealed in a 10 mTorr vacuum at about $55^oC$ for various lengths of time, during which time the room temperature sheet resistance decreased by up to ten percent. The film thickness was monitored during the ebeam evaporation by a quartz crystal monitor near the sample, and confirmed by an edge-on SEM image, or by X-ray reflectivity. Typical films used in our present study were 200 - 300 $\AA$. X-ray scattering yielded a broad peak characteristic of an amorphous film. In table \ref{table} we list the properties of the films that we discuss in this paper; detailed information is given in the next several sections. The films were mounted at the bottom of an Oxford 400 dilution refrigerator. The linear sheet resistance was measured using standard four-point lockin techniques at frequencies from 3 - 17 Hz and excitation currents 0.1 - 5 nA, depending on the resistance range; the highest resistance sample was measured at dc.

\begin{table}[h]
\centering
\caption{ \footnotesize \setlength{\baselineskip}{0.8\baselineskip} Parameters for the samples discussed in this paper.  We name the samples as weak, intermediate and strong according to their insulating behavior.  $T_{c0}$ is the mean-field transition temperature, $R^*_n$ is the sheet resistance of the samples at $T = 2 T_{c0}$, $H_{c2}(0)$ and $\xi(0)$ are calculated using the slope near $T_{c0}$ and do not reflect low-temperatures corrections due to disorder. $\langle H_c\rangle$ and $R_c$ are the critical field and resistance of the SIT.  For more discussion on the parameters of the films see text.}
    \begin{tabular}{|c|c|c|c|}
      \hline
     Quantity\hspace{4mm}  & \hspace{5mm}Weak\hspace{5mm} &Intermediate & \hspace{4mm}Strong\hspace{4mm} \\
     \hline
      \hline
         Film thickness $ [\AA]$ &300 & 200 & 300 \\
            \hline
$T_{c0}$ $ [K]$ &3.35 & 2.9 & 2.7 \\
            \hline
 $\xi(0)$ $ [\AA]$ &50 & 58 & 60 \\
            \hline
$R^*_n$  [k$\Omega]$ & 2.6 & 5.0 & 6.8 \\
            \hline
$H_{c2}(0)$ [T]  & 13 & 9.5 & 9 \\
     \hline$ \langle H_c\rangle$ [T]  & 11.2 & 9.0 & 3.4 \\
     \hline
     $R_c$  [k$\Omega]$ & 4.05 & 4.5 & 7.0 \\
     \hline
    \end{tabular}
    \label{table}
  \end{table}

\section{The Superconducting Transition and $H_{c2}$}
\label{sechc2}

\begin{figure}[ht]
\centering
\includegraphics[width=1.0\columnwidth]{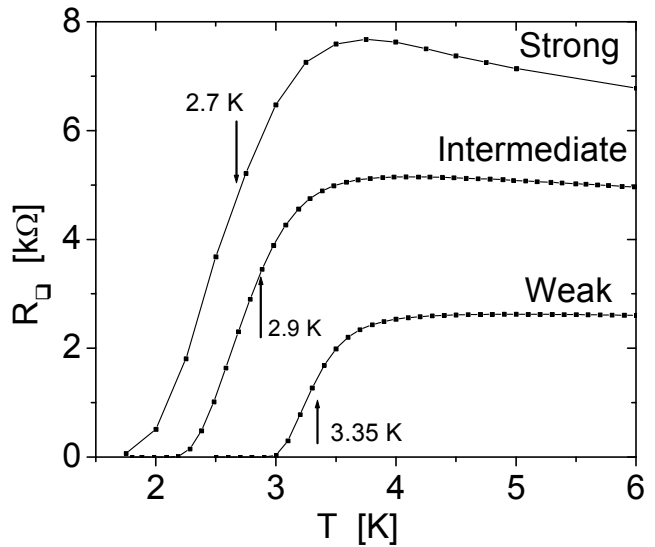}
\caption{ \footnotesize \setlength{\baselineskip}{0.8\baselineskip} Zero-field superconducting transitions for three samples. The labels denote the strength of the insulating phase, discussed later in the text. The mean field transition temperatures $T_{c0}$ are indicated by the vertical arrows.}
\label{Tco}
\end{figure}

To understand the SIT behavior of the films at low temperatures and high magnetic fields we first need to understand their transition to the superconducting state, and estimate the mean-field upper critical field as projected from near $T_{c0}$. Fig.~\ref{Tco} shows the zero-field transitions for three samples. We classify these samples as ``weak," ``intermediate" and ``strong" in terms of their insulating behavior to be discussed later. These three samples show increasing normal state resistance which we will take as a measure of increasing disorder. \cite{mpaf}  For temperatures above $T_{c0}$ in the range of 8 -- 30 K (not shown), the sheet resistance varies as $R_{n}(T) = R_{0} + \Delta R_{\Box}(T)$, with $\Delta R_{\Box}(T) \sim T^{-\alpha}$.  We find that $\alpha \approx$ 0.085, 0.22 and 0.41 for the three samples, and we take this as the phenomenological normal state $R_n(T)$ at these temperatures. At lower temperatures, where the inelastic scattering time exceeds the film thickness, we expect the normal state to be well-described by the usual weak  localization $log(1/T)$ correction as do similar, non-superconducting InOx films.  \cite{ovadyahu}

Before we turn to the evaluation of $H_{c2}$ we need to clarify its definition in the presence of strong phase fluctuations.  In a mean-field approximation, $H_{c2}$ marks the magnetic field for which the superconducting transition, $T_c(H)$ occurs, and $H_{c2}(0)$  is the zero temperature limit. This field is determined by order-parameter amplitude considerations such that above it all pairs are broken and the Abrikosov lattice of vortices has disappeared. \cite{tinkham} However, going beyond the mean-field approximation to include strong fluctuations, and in particular for two-dimensional or quasi-two-dimensional superconductors with low superfluid density, melting occurs at fields that can be well below the mean-field $H_{c2}$, marking the {\it true} superconducting transition that is dominated by phase fluctuations.\cite{kt,yazdani1}  Typically for such systems, amplitude fluctuations will be very weak above melting and therefore the amplitude of the order parameter can be considered to be almost constant, allowing for the local definition of vortices and vortex liquid.  With increasing magnetic field, amplitude fluctuations increase and the system crosses over to a system dominated by fermions.  The notion of an upper critical field as a unique magnetic field above which all aspects of superconductivity are gone is therefore lost.  However, one can still discuss the mean-field value of $H_{c2}$ as a scale for pair breaking, especially at low fields.  In fact, it was shown by Urbach {\it et al.} \cite{urbach} using specific heat measurements of thin MoGe films that most of the entropy of the superconducting state is recovered at the mean field $H_{c2}$.  Thus, in what follows we estimate $H_{c2}$ in the mean-field sense, projecting its low temperature values from the behavior near $T_{c0}$.

\begin{figure}[ht]
\centering
\includegraphics[width=0.9\columnwidth]{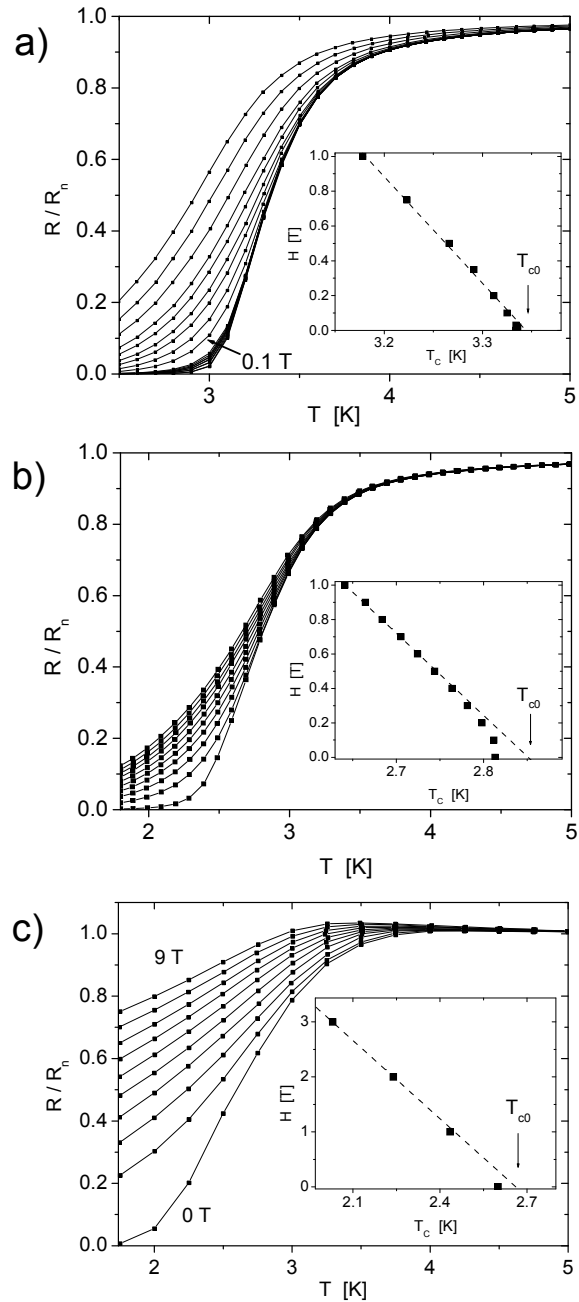}
\caption{ \footnotesize \setlength{\baselineskip}{0.85\baselineskip} Low field superconducting transitions: (a) the weak sample. The overlapping curves below 0.1 T are at 0.005, 0.01, 0.015, 0.02, 0.025 and 0.03 T, while those above are at 0.2, 0.35, 0.5, 0.75, 1, 1.5, 2 and 2.5 T; (b) the intermediate insulator in steps of 0.1 T; (c) the strong insulator in steps of 1 T. The insets show $H_{c2}(T)$ taken at $R/R_n = 0.5$. The measured $H_{c2}(T)$ dives down below (a) 0.05 T, (b) 0.2 T and (c) 1.5 T, while the dashed lines extrapolate to $T_{c0}$.}
\label{hc2}
\end{figure}

To evaluate $T_{c0}$ and the upper critical field $H_{c2}$ we analyze the superconducting transitions at low fields, shown for example in Fig.~\ref{hc2} for the three samples; all plots have been normalized by their respective $R_n(T)$. At the very lowest fields the resistance curves almost fall on top of each other through most of the transition. This is most pronounced in the weak insulator of Fig.~\ref{hc2}a, where the six curves with $H \leq 0.04$ T are practically indistinguishable. This kind of behavior could indicate that a pair-breaking length $\ell_i$ shorter than the magnetic length $\ell_H = (\Phi_0/2\pi H)^{1/2}$ cuts off the superconducting fluctuations. \cite{maekawa1} With increasing field, the magnetic length decreases and when it is the shorter of the two lengths, the resistive transitions start to separate in a manner expected for a superconductor with fluctuating amplitude. We note that this procedure is different from the one suggested by Hebard and Paalanen \cite{hebard3} which adds a pair-breaking time linearly with the field.  Their procedure results in a reduction of the transition temperature, but it retains a linear dependence of the field on temperature near $T_{c0}$,  in contrast to our experimental findings (see the insets to Fig.\ref{hc2}). As superconducting fluctuations start to affect the normal state resistance, the resistance starts to drop, at first in a more mean-field-like manner. Deeper into the transition the resistance curves broaden substantially and tail exponentially to zero resistance as expected for a true two-dimensional superconductor. To estimate $H_{c2}(T)$ we first plot $R(T)/R_n(T)$ vs. $T$ as in Fig.~\ref{hc2}. The upper part of the transitions, at fields greater than the overlapping fields, displays parallel curves with increasing magnetic field that allows us to extract the mean-field transition temperatures $T_c(H)$ by taking horizontal cuts through the data. The insets to Fig.~\ref{hc2} show the resulting temperatures taken at $R/R_n = 0.5$. The linear field regime extrapolates on the temperature axis to $T_{c0}$, while the slope, multiplied by $0.69 T_{c0}$ gives an estimate of $H_{c2}(0)$ as would be calculated by WHH theory.  \cite{whh}  From the slopes of the insets we calculate zero-temperature coherence lengths $\xi(0) \approx$ 50 - 60$\AA$, very reasonable for these highly disordered films.

While the choice of $R/R_n = 0.5$ to mark the mean field transition is somewhat arbitrary, we do not expect the actual value to be far from this determination.  To arrive at the above criterion we first fit the zero-field transition to an Aslamazov-Larkin (AL) form. \cite{tinkham} We then note that cuts that are close to $R/R_n = 0.5$ have their linear part of $T_c(H)$ extrapolating to the transition determined by the AL fit.    Since these films are strongly fluctuating, the notion of an exact mean-field transition temperature $T_{c0}(H)$ is lost, like $H_{c2}$, and we are left with an estimate of the energy scale for superconductivity. The uncertainty in $T_{c0}(H)$ may translate into an uncertainty in our estimate of $H_{c2}$, but this is relatively unimportant compared to the overall field scale of our measurement, and in particular compared to the field scale of the magnetoresistance peak that we shall discuss in section \ref{TheSIT}. We shall return to this point later. 

The above procedure gives the values for $H_{c2}(0)$ of 13, 9.5 and 9  T for the weak, intermediate and strong insulating samples respectively, accurate to within $\sim 10 \%$ based on the estimates of $T_{c0}(H)$. However, we expect that the actual $H_{c2}(0)$ will be higher due to disorder-induced pair-breaking and to increased Coulomb interactions which affect the density of states.  Following Maekawa {\it et al.} \cite{maekawa1} we first estimate the elastic disorder parameter $\lambda = \hbar/2\pi \epsilon_F\tau$.  We find that $\lambda$ is in the range of 0.05 to 0.1 for our films, where $\tau$ is the elastic relaxation time. Using this parameter we estimate an increase of the slope (and hence $H_{c2}(0)$) of about 15 - 25$\%$. \cite{maekawa1} We also expect a further increase in slope at lower temperatures due to enhanced Coulomb interaction in these low carrier-density films. \cite{maekawa1}  While we cannot estimate this enhancement, previous measurements on indium-oxide films by Hebard and Paalanen \cite{hebard3} clearly show an enhancement of $H_{c2}(0)$. The resulting $H_{c2}(0)$ for all three samples is therefore expected to be in the vicinity of 13 - 14 T.   These estimates are important in order to understand the Bose-insulating phase and its crossover to a Fermi system which will be discussed in the section \ref{insu}.

The above procedure also allows us to calculate the inelastic length or phase-breaking time that cuts off the magnetic length close to the transition. Below this length $\ell_i$, the $H_{c2}(T)$ curve dives down as a function of temperature instead of continuing linearly to $T_{c0}$ at zero field, as shown in the insets to Fig.~\ref{hc2}. This deviation from a linear $H_{c2}$ curve occurs at fields of $H_i = 0.05$ T , $H_i = 0.2$ T, and $H_i = 1.5$ T for the the weak, intermediate and strong insulating samples respectively, with corresponding lengths $\ell_i \approx 900 \AA$, $\ell_i \approx 400 \AA$, and $\ell_i \approx 150 \AA$. Thus we see that $\ell_i$ is the longest for the least disordered sample. We note that this treatment {\it assumes} the existence of a pair-breaking length, $\ell_i$. While agreeing well with our experimental finding, it is nevertheless clear that more theoretical work is needed to justify such an approach.

\section{Intermediate Fields}

\begin{figure}[ht]
\centering
\includegraphics[width=0.9\columnwidth]{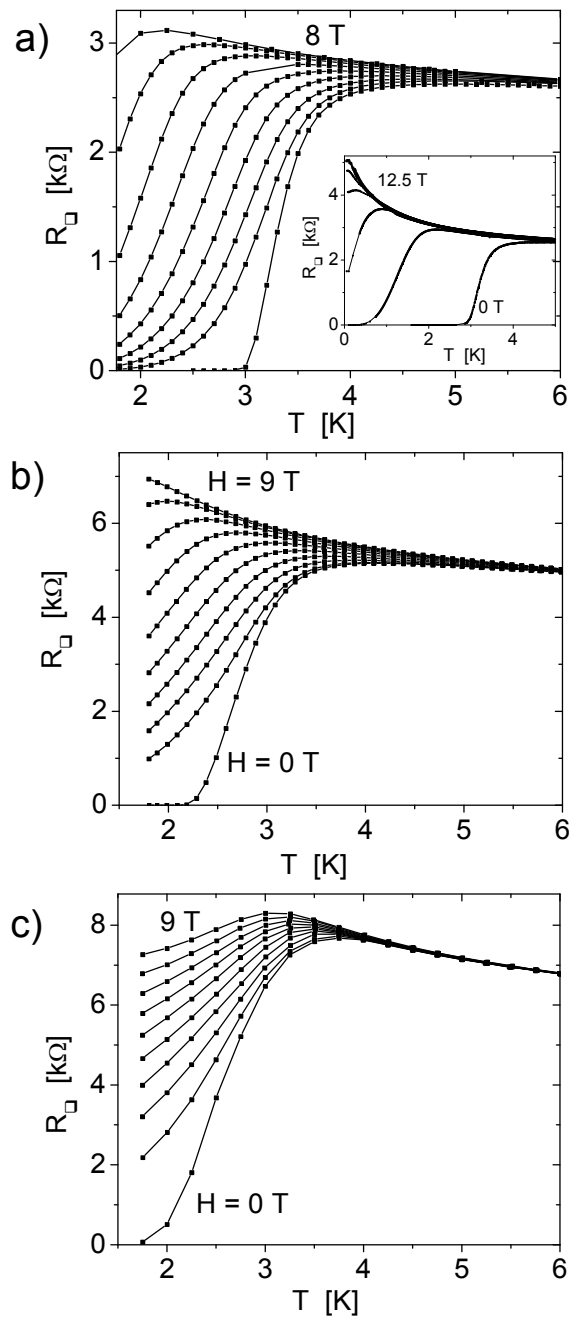}
\caption{ \footnotesize \setlength{\baselineskip}{0.8\baselineskip} Resistance envelopes for the (a) weak, (b) intermediate, and (c) strong insulators. The fields are in steps of 1 T, beginning with zero-field. The inset to (a) shows the envelope of the weak sample extending to 100 mK and 12.5 T; the fields are 0, 8.0, 10.5, 11.2, 11.5, 12.0 and 12.5 T. } 
\label{Env}
\end{figure}

Measuring the resistance at higher fields, we observe the behavior shown in Fig.~\ref{Env}, where the data is dominated by the presence of a resistance envelope. The curves at progressively higher fields follow the envelope to lower temperatures before they depart toward zero. Figures \ref{Env}a and \ref{Env}b show the weak and intermediate insulators, respectively, and are qualitatively very similar up to 8 T.  The envelope for the strong insulator,  shown in Fig.~\ref{Env}c, is less developed at these fields. The inset to Fig.~\ref{Env}a shows the envelope for the weak insulator extending to 12.5 T and 100 mK, with all curves following to within $\sim 2\%$.   The curve appear to saturate below $\sim$ 100 mK, similar to our previous results on MoGe.\cite{mk1} We checked the linearity of the $I-V$ to make sure that the saturation was not due to heating, and the dilution refrigerators' cooling (two different ones, one in our lab and one at the NHMFL) to make sure that the saturation was not due to a lack of further cooling power.  At temperatures much lower than $T_{c0}$ the three samples presented in this paper show very different insulating tendencies. While the weak insulator sample shows an envelope consistent with a logarithmic increase, the strong insulator sample shows an exponential increase of the resistance with decreasing temperatures. We shall discuss these different tendencies in section \ref{insu}.

The envelope patterns presented here are qualitatively different from other reports on InOx. \cite{hebard1, gantmakher2, valles2} Physically, the pattern indicates that most of the transition is dominated by amplitude fluctuations, with a final phase-dominated SIT at low temperature. By contrast, early work \cite{hebard1} on InOx showed curves that appear to splay off from a common temperature, typically with wider transitions (see figure 1 in Hebard and Paalanen \cite{hebard1}). That pattern is more consistent with a granular, phase-dominated system: the initial transition from a common temperature results from the  superconducting transition within each grain, even as the overall behavior is dominated by the Josephson phase-coupling between grains; increasing the field effectively reduces the tunnelling between grains and therefore suppresses the overall superconductivity of the sample, manifest by a larger transition width. Gantmakher {\it et al.} \cite{gantmakher2} studied homogenous InOx films and did not observe a splaying of the curves at a common temperature, but neither did they observe a resistance envelope. Their data shows a critical point and a negative magnetoresistance at high fields similar to our own, as will be discussed later. In a different experiment, Gantmakher {\it et al.} \cite{gantmakher} reported results on an amplitude-dominated Nd$_{2-x}$Ce$_x$CuO$_{4+y}$ (NdCeCuO) layered film, with a pattern similar to Fig.~\ref{Env}a, though any envelope behavior is not clear, and not discussed. Hsu and Valles \cite{valles2} observed a resistance envelope in quench-condensed Pb films, but with much higher $R_n(T)$ and a granular morphology. They focused on the shift in the superconducting transition to lower temperature and did not comment on the envelope.

\section{The Superconductor-Insulator transition} 
\label{TheSIT}

\begin{figure}[ht]
\centering
\includegraphics[width=0.85\columnwidth]{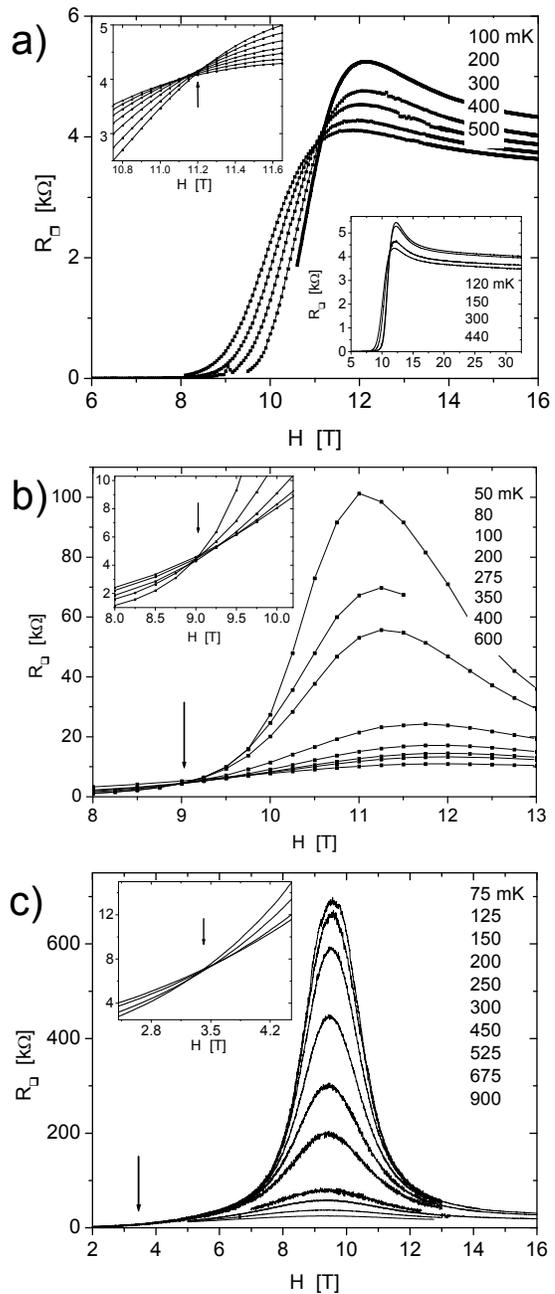}
\caption{ \footnotesize \setlength{\baselineskip}{0.8\baselineskip}  Resistance isotherms in a perpendicular magnetic field: (a) the weak, (b) intermediate and (c) strong insulators. The temperatures of the isotherms (at the peaks) are listed on the graphs. The upper insets show the crossing point in the range 100 - 500 mK. $\langle H_c\rangle$ for each film is indicated by the arrows. The lower inset to (a) shows four isotherms measured to 32.5 T at the National High Magnetic Field Laboratory. }
\label{rvsh}
\end{figure}

In the high field and low temperature regime the system shows the signature of a superconductor-insulator transition. \cite{hebard,mk1,goldman1} In Fig.~\ref{rvsh} we show isotherms of the resistance  at high magnetic fields for the three films. There are several interesting features common to all three sets of data. First, the isotherms go through a temperature-independent crossing point, a signature of a zero temperature quantum phase transition. This is most apparent in Figure \ref{rvsh}a, while in Figures \ref{rvsh}b and \ref{rvsh}c it is less so because of the expanded vertical scale. A closer inspection of the crossings, shown in the upper insets to Figure \ref{rvsh}, indicates that the transition broadens at low temperatures, $\Delta H_c /\langle H_c \rangle \sim 1\%$, a feature previously discussed by Mason and Kapitulnik.\cite{mk1} Nevertheless, we can use $\langle H_c \rangle$ to scale the data using the usual one parameter scaling form: $R(H,T) = R_c \mathcal{F}\left(\frac{H - \langle H_c\rangle}{T^{1/z\nu}}\right)$. \cite{hebard1,kap2}  The above fit gives $z\nu \simeq 1.3$, in agreement with other measurements on MoGe \cite{kap1,mk1} and InOx. \cite{hebard1}  The crossing point marks the transition from a regime where the resistance decreases with decreasing temperature to a one where it increases with decreasing temperature. This latter regime is insulating, at fields that are clearly lower than the zero-temperature mean-field $H_{c2}(0)$ estimated above.  The resistances  $R_c$ at the transition are 4.05, 4.5, and 7.0 k$\Omega/\Box$ for the three samples, showing that $R_c$ tends to increase with increasing disorder.  The fact that $R_c$ for the most disordered sample exceeds the quantum of resistance, $R_Q \equiv h/(2e)^2 \approx$ 6.5 k$\Omega$ may indicate again that even for strongly disordered samples, the self-duality between pairs and vortices which results in $R_c = R_Q$ is not satisfied, and moreover, the critical resistance is non-universal. The critical fields are 11.2, 9.0 and 3.4 T respectively, showing a substantial decrease in $\langle H_c\rangle$ with increasing disorder. \cite{mpaf}  

\section{The Insulating Phase}
\label{insu}

Upon increasing the magnetic field we note that the isotherms reach a maximum and then start to decrease.  The maximum for all samples is very close to the mean-field $H_{c2}(0)$ which we previously estimated from the slopes near $T_{c0}$. This is expected if we attribute the peak to the breaking of pairs and the crossover of the system from being Bose-particle dominated to being Fermi-particle dominated. 

As a specific example let us consider Fig.~\ref{rvsh}a. The isotherms cross at $H_c \sim 11.2$ T and go through a resistance peak near 12.5 T. This field coincides with the maximum field along the envelope in the inset to Fig.~\ref{Env}a; data taken at higher fields have resistances below the envelope. Additionally, the resistance along the envelope exceeds the expected normal state resistance such that at the peak the system is clearly not in its normal state. The peak position is not strictly temperature independent but shifts to higher fields as the temperature is lowered, which may explain the small width of the envelope. The magnetoresistance with respect to the peak remains negative all the way to 32.5 T, the upper limit of the measurement as shown in the lower inset to Fig.~\ref{rvsh}a. While the isotherms appear to gradually level at high fields, it does not appear based on the available data range that it actually saturates at the highest accessible field. Let us point out that the resistance at lowest temperature (120 mK) at 32.5 T is almost a factor of 1.7 higher than the normal resistance at zero field, just above the transition. 

To estimate the expected classical normal state resistance (at high field, in the absence of superconductivity) we use a simple ``Kohler's rule" for the scaling of the magnetoresistance. \cite{ziman} This is justified because Kohler's rule is valid for any metallic system as long as there is only one dominant scattering time; InOx is a simple amorphous metal that has been studied in the past and found to satisfy this condition. \cite{ovadyahu} The magnetoresistance scales as $\Delta \rho/\rho_n = f(\omega_c\tau)$, where $\rho_n$ is the normal state resistivity, $\omega_c = (eH/mc)$ is the cyclotron frequency, $\tau$ is the scattering time, and $f(x)$ is a scaling function which for most fields in the range of interest should be approximately $f(x) \sim Ax^2$ with $A$ a constant of order unity. \cite{kapiza} For free electrons the above relation can also be written as $\Delta \rho/\rho_n = f(\rho_ H/\rho_n)$, where $\rho_H$ is the Hall resistivity. Using the measured carrier density $n \sim 10^{20}$ carriers/cm$^3$ we expect that classically $\Delta \rho/\rho_n \sim 10^{-3}$, much lower than the observed magnetoresistance by a factor of almost 10$^3$. Thus it appears that even at the highest field the film shows vestiges of pairing.

The insulating state above the crossing point becomes even more dramatic in the films with higher $R_n$. At low temperature and moderately high field these films reveal a new, extremely strong tendency towards the insulating phase, shown in Figures \ref{rvsh}b and \ref{rvsh}c. As we have noted, the resistance at the crossing point of both plots is comparable to that of the first film, though the position of the crossing point shifts to lower field as the insulator strength increases. On the high field side of the peak the isotherms all decay to resistances $\leq$ 20 k$\Omega/\Box$, above the normal resistance $R_n$ even at the highest accessible fields, again indicating incipient superconductivity much above the mean field $H_{c2}$.

\begin{figure}[ht]
\centering
\includegraphics[width=0.9\columnwidth]{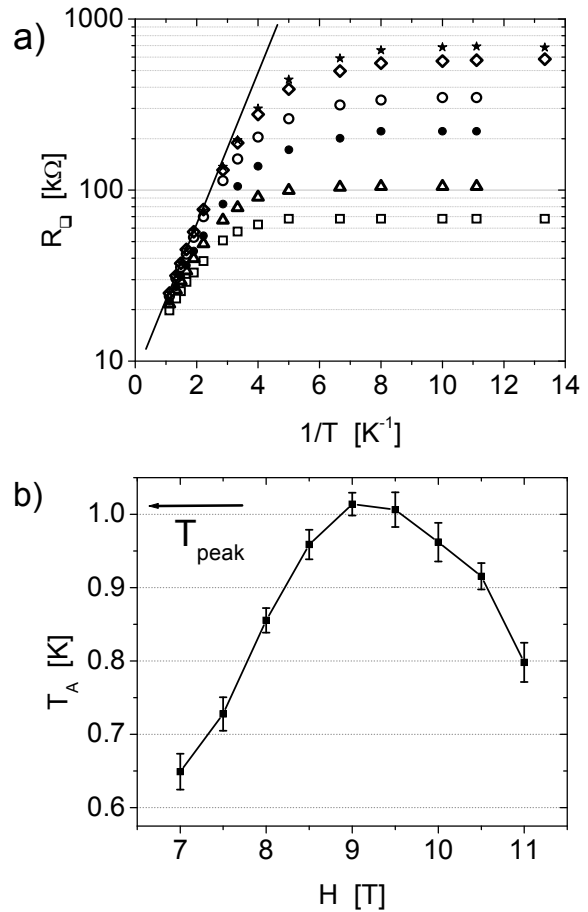}
\caption{ \footnotesize \setlength{\baselineskip}{0.8\baselineskip}  (a) Arrhenius plot of the resistance of the strong insulator, following $R\sim e^{(T_A/T)}$. The straight line indicates the activation regime. The fields are: $\Box$ 7.0 T, $\triangle$ 7.5 T, $\circ$ 8.5 T, $\diamond$ 9.0 T, $\star$ 9.5 T, $\bullet$ 11.0 T.  (b) The activation energy extracted from (a) as a function of field. The peak temperature gives a measure of the insulator strength. }
\label{Active}
\end{figure}

\begin{figure}[ht]
\centering
\includegraphics[width=0.9\columnwidth]{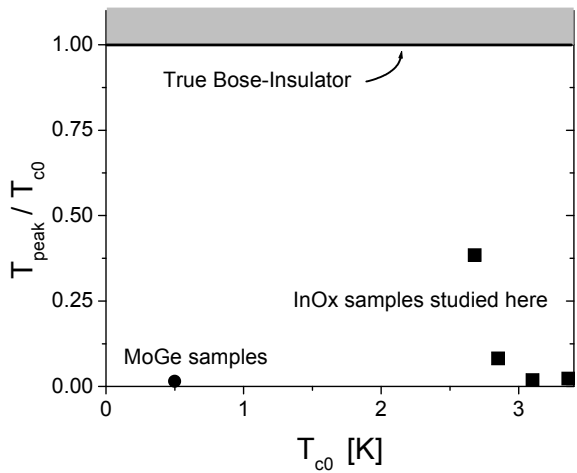}
\caption{ \footnotesize \setlength{\baselineskip}{0.8\baselineskip} A phase-space plot. The relative strength of the insulating and superconducting phases is plotted as the ratio $T_{peak}/T_{c0}$ vs $T_{c0}$. We included another InOx film, as well as an analysis of the MoGe data of Mason {\it et al}.\cite{mk1} }
\label{PhaseD}
\end{figure}

The increase in resistance in these stronger insulators is several orders of magnitude before the resistance starts to saturate at very low temperatures. Taking fixed-field cuts through the isotherms, \cite{sambandamurthy1} we can study this increase in resistance by constructing an Arrhenius plot of $R_\Box \sim e^{T_{A}/T}$ with a characteristic activation energy $k_BT_A$.  We plot such a curve in Fig.~\ref{Active}a for the strongest insulator; the other films produce similar, if less pronounced, results, as might be expected. Indeed at the higher temperatures activation is recovered with higher activation energy corresponding to higher field, until we get to the peak field, shown in Fig.~\ref{Active}b. At that point the activation energy is $\sim$ 1 K for the strongest insulator. This provides a measure of the characteristic energy scale of the insulator. We note that while this procedure is very well defined for the strong and intermediate insulators, it is less well-defined for the weak insulator where the temperature dependence of its envelope is more logarithmic in shape.  However, to have a unifed characterization for all samples we enforce an activated behavior and thus  extract the peak activation temperatures for all of the films, and compare them with $T_{c0}$, the energy scale of the superconductor. Fig.~\ref{PhaseD} shows a phase space plot that illustrates this relationship. We have included points derived from other InOx films and from  previous measurements of MoGe films \cite{mk1} that show the same physics. Even the strongest insulator in our sample set only falls in the middle of the graph. We identify the line $T_{peak}/T_{c0} = 1$ with a true Bose-insulator, where we expect all of the electrons in the system to enter as localized pairs. For such a sample we also expect that no saturation of the resistance at the lowest temperatures will be found and a true SIT will be recovered. Scaling about the crossing point for such a system is then expected to be dominated by quantum percolation as was discussed by Kapitulnik {\it et al.} \cite{kap1}

\section{Discussion and Conclusions}

The picture that emerges for the InOx films studied here is of a superconducting system with regimes that are dominated by amplitude fluctuations and regimes that are dominated by phase fluctuations. At low fields, near $T_{c0}$, amplitude fluctuations dominate the behavior of the superconducting transition, giving rise to a broadening of the transitions and an overall resistance envelope. Analyzing the transitions to extract the upper critical field, we find that $H_{c2}(T)$ approaches $T_{c0}$ linearly, then deviates toward $H=0$ with infinite slope, suggesting that the magnetic length is ultimately cut off by a shorter inelastic pair-breaking length, the character of which is presently unknown.

For higher fields and lower temperatures we observe a phase-dominated superconductor-to-insulator quantum phase transition, \cite{mpaf} indicated by the crossing point in the isotherms.  Above the crossing point, a reduction in temperature strengthens the localization and the resistance appears activated with a characteristic energy that is always lower than $k_BT_{c0}$, but approaches this value as the strength of the insulator increases. 

The magnetic field appears to be doing two competing things: increasing the field drives the system away from the SIT deeper in to the Bose-insulating phase, while at the same time it directly depairs the Cooper pairs. This latter effect is clearly an amplitude effect and is strongly identified with the peak in the isotherms and the envelope in the resistance. At that point, enough pairs have been broken so that the resistance is dominated by the scattering of the free fermions.  A similar idea was suggested by Hebard {\it et al.} \cite{hebard2} based on a crossing point in the Hall resistivity near the peak.  Our estimate of $H_{c2}(0)$ give values in the vicinity of the peak, reinforcing this interpretation.

Finally, above the peak, the resistance decays very slowly out to high fields, remaining at values above its expected normal state. We suggest that this results from a non-negligible residual density of pairs beyond the peak. We are left with the hypothesis that as the magnetic field suppresses the pair-amplitude {\it very} slowly out to high fields, the system retains a vestige of superconductivity at fields well above $H_{c2}$. This conjecture may have broader application to the high-$T_c$ superconductors: a remarkably similar envelope behavior was observed \cite{ando} in underdoped $\rm La_{2-x}Sr_xCuO_4$, where the envelope persisted to 50 T. As such, our results may contribute to a new understanding of the ``normal state" found in the cuprates for high fields. This possiblity is further discussed in a separate publication. \cite{myles1}

In summary, we present studies of amorphous InOx films showing that their behavior at low fields is dominated by amplitude fluctuations. The superconductor-insulator transition in these films occurs at $R_\Box$ of order of the quantum value. The insulating phase varies in strength revealing a strong magnetoresistance peak that marks the crossover from a Bose-insulating to a high-field fermion-dominated behavior. The existence of a large magnetoresistance at fields well above the peak suggests that superconducting pairs persist in these systems at fields well above the mean field $H_{c2}(0)$.\\

\noindent {\bf Acknowledgments:} We thank Yoichi Ando, Greg Boebinger, Steven Kivelson, and Sadamichi Maekawa for very helpful discussions, and  to Tim Murphy and Eric Palm at NHMFL in Florida for their help with the experiment at the magnet lab, funded by NSF and the State of Florida. Work supported by the National Science Foundation Grant:  NSF-DMR-0119027.

\end{document}